\documentclass[aps,preprint]{revtex4}%
\usepackage{amsfonts}
\usepackage{amsmath}
\usepackage{amssymb}
\usepackage{graphicx}%
\providecommand{\U}[1]{\protect\rule{.1in}{.1in}}
\begin{document}
\title{Correlation strengths in hybrid networks}
\author{Li-Yi Hsu}
\affiliation{Department of Physics, Chung Yuan Christian University, Chungli 32081, Taiwan}
\affiliation{Physics Division, National Center for Theoretical Sciences, Taipei 106319, Taiwan}

\begin{abstract}
In a generic hybrid network, classical, quantum, and no-signaling sources emit
local hidden variables, stabilizer states, and no-signaling systems,
respectively. We investigate the maximal correlation strength as the
non-classical feature in this network. Given the associated fully-quantum
network of a hybrid network, we exploit the stabilizing operators of the
distributed quantum state to construct segmented Bell operators and the Bell
inequalities tailored to the state. We derive the upper bounds of the maximal
correlation strengths in the associated full-classical, full-quantum, and
fully-no-signaling networks as the benchmarks. Our study shows that the
achievable correlation strength depends on the number of type-$A$ measurements
and that of nonlocal sources. We also introduce the $t$-nonlocality criteria,
indicating that the achievable maximal correlation strength cannot modeled by
the network with at least $t$ observers with local hidden variables performing
type-$A$ measurements.

\end{abstract}
\volumeyear{year}
\volumenumber{number}
\issuenumber{number}
\eid{identifier}
\startpage{1}
\endpage{2}
\maketitle

\textit{Introduction }With advancements in quantum information science, Bell
nonlocality represents a key resource for various cryptographic tasks
\cite{3,4,5}. In standard Bell tests on correlation statistics, a single
source emits a state of physical systems to distant observers. Upon receiving
these systems, each observer randomly chooses the setting of the local
measurement as the input. After the local measurement, each observer obtains
the measurement outcome as output. Bell theorem states that the input-output
correlations based on classical models cannot replicate the predictions of
quantum theory. Since local realism imposes constraints on the classical
systems, classical correlation strength can be simulated using some local
hidden variables (LHV) and obey any Bell inequality \cite{2}. On the other
hand, quantum theory allows for stronger correlations that can violate Bell
inequalities, and the hypothetical systems governed by the no-signaling
principle can achieve the algebraic maximal violation beyond the quantum
region \cite{66,77}.

As quantum communication develops rapidly, growing research interest focuses
on non-classical correlations within networks involving multiple independent
sources. In most literature, the study explores the achievable correlation
strength in the quantum network with all independent sources emitting quantum
states \cite{review}. On the other hand, a hybrid network involves classical,
quantum, and no-signaling sources emitting LHVs, quantum states, and
no-signaling systems, respectively. Classical sources are local, while quantum
and no-signaling sources are nonlocal. Hence the achievable correlation
strength in hybrid networks becomes more sophisticated than that of quantum
networks. For example, in the two-source scenario, the degree of violation of
the bilocal inequality using two quantum sources is equivalent to that using
one classical source and one no-signaling source \cite{fnn}. Therefore, the
non-classical features can exist in the hybrid networks even when classical
local sources are present. There are alternative approaches to tackle the
complicated correlation in networks. For instance, the inequalities for
certifying the non-classicality in three-source star networks are reported
using the inflation technique \cite{infla}, and various hierarchies of network
nonlocality levels have been investigated in \cite{hier}.

In this work, we explore the achievable correlation strength in the generic
network $\mathcal{N}^{(n,M)}$ which involves $n$ sources and $M$\ observers.
The $i$-th classical, quantum, or no-signaling\ source $e_{i}^{(C)}$,
$e_{i}^{(Q)}$ or $e_{i}^{(NS)}$ each emits $N_{i}$ classical, quantum or
no-signaling\ systems indexed as $(i,1)$, \ldots, and $(i,N_{i})$,
respectively. There are $N=%
%TCIMACRO{\dsum \nolimits_{i=1}^{n}}%
%BeginExpansion
{\displaystyle\sum\nolimits_{i=1}^{n}}
%EndExpansion
N_{i}$ systems distributed in $\mathcal{N}^{(n,M)}$. Specifically, the fully
classical, fully quantum fully no-signaling networks are denoted as
$\mathcal{N}_{C}^{(n,M)}$, $\mathcal{N}_{Q}^{(n,M)}$ and $\mathcal{N}%
_{NS}^{(n,M)}$, where all sources in $\mathcal{N}^{(n,M)}$ emitting classical,
quantum and no-sginaling systems, respectively. In the following, the observer
receives the system or particle with the same index in any configuration. The
notation $(i,j)\rightarrow\mathcal{K}$ indicates that the $j$-th particle
emitted from the source $i$ is sent to the observer $\mathcal{K}$. Given a
generic hybrid network $\mathcal{N}^{(n,M)}$, we can construct the Bell
operators based on $\mathcal{N}_{Q}^{(n,M)}$ with the quantum sources emitting
highly entangled stabilizer states. The stabilizing operators are crucial for
designing Bell operators with appropriate measurement assignments. In
addition, the maximal correlation strength of $\mathcal{N}_{C}^{(n,M)}$ and
$\mathcal{N}_{NS}^{(n,M)}$serve as the local upper bounds and the algebraic
maximum of the proposed Bell inequalities tailored for $\mathcal{N}^{(n,M)}$,
respectively. Later the notation $\mathcal{N}^{(n,M)}$ will be alternatively
denoted as $\mathcal{N}_{H}^{\left\langle u,v,w\right\rangle }$ for the
further clarification of a hybrid network.

\textit{Bell operators} Let the quantum\ sources $e_{i}^{(Q)}$ in
$\mathcal{N}_{Q}^{(n,M)}$ emit the $N_{i}$-qubit stabilizer state $\left\vert
\psi_{i}\right\rangle $, resulting in the product state $\left\vert
\Psi\right\rangle =%
%TCIMACRO{\dprod \nolimits_{i=1}^{n}}%
%BeginExpansion
{\displaystyle\prod\nolimits_{i=1}^{n}}
%EndExpansion
\left\vert \psi_{i}\right\rangle $ distributed within the fully-quantum
network. The observers $\mathcal{A}_{i}$ and $\mathcal{B}_{j}$ receive
$n_{A,i}$ and $n_{B,j}$ qubits, where $i=1,\ldots,M_{A}$ and $j=1,\ldots
,M_{B}$, with $M_{A}+M_{B}=M$ and $%
%TCIMACRO{\dsum \nolimits_{i=1}^{M_{A}}}%
%BeginExpansion
{\displaystyle\sum\nolimits_{i=1}^{M_{A}}}
%EndExpansion
n_{A,i}+%
%TCIMACRO{\dsum \nolimits_{j=1}^{M_{B}}}%
%BeginExpansion
{\displaystyle\sum\nolimits_{j=1}^{M_{B}}}
%EndExpansion
n_{B,j}=N$. Unless otherwise specified, let $n_{A,i}=1$ and $n_{B,j}\geq2$.
For a given $M_{A}$-bit string $l$ $=l_{1}l_{2}...l_{M_{A}}$, there are
associated $n_{B,j}$-bit binary\ string $y_{j(l)}=y_{j(l),1}\ldots
y_{j(l),n_{B,j}}$, local operator $g_{l}^{(i)}$and global stabilizing operator
$g_{l}=%
%TCIMACRO{\dprod \nolimits_{i=1}^{n}}%
%BeginExpansion
{\displaystyle\prod\nolimits_{i=1}^{n}}
%EndExpansion
g_{l}^{(i)}$ such that $g_{l}^{(i)}\left\vert \psi_{i}\right\rangle
=\left\vert \psi_{i}\right\rangle $ and hence $g_{l}\left\vert \Psi
\right\rangle =\left\vert \Psi\right\rangle $. Alternatively, we express
$g_{l}$ as
\begin{equation}
g_{l}=%
%TCIMACRO{\dprod \nolimits_{i=1}^{M_{A}}}%
%BeginExpansion
{\displaystyle\prod\nolimits_{i=1}^{M_{A}}}
%EndExpansion
O_{A,l}^{(i)}%
%TCIMACRO{\dprod \nolimits_{j=1}^{M_{B}}}%
%BeginExpansion
{\displaystyle\prod\nolimits_{j=1}^{M_{B}}}
%EndExpansion
O_{B,l}^{(j)}, \label{stabilizer}%
\end{equation}
where $O_{A,l}^{(i)}$ is a Pauli observable, and $O_{B,l}^{(j)}$ is an
$n_{B,j}$-fold tensor products of Pauli observables. Exactly one of the
following three conditions must hold:
\begin{equation}
(i)\text{ }O_{A,l}^{(i)}\in\{X,Y\},(ii)\text{ }O_{A,l}^{(i)}\in
\{Y,Z\},(iii)\text{ }O_{A,l}^{(i)}\in\{X,Z\}, \label{criteria}%
\end{equation}
where $X$, $Y$, and $Z$ denote the Pauli observables $\sigma_{x}$, $\sigma
_{y}$, and $\sigma_{z}$, respectively.\ Denote the observer set as
$\mathcal{O}_{\mathcal{A}}=\{\mathcal{A}_{1},\cdots,\mathcal{A}_{M_{A}}\}$ and
$\mathcal{O}_{\mathcal{B}}=\{\mathcal{B}_{1},\cdots,\mathcal{B}_{M_{B}}\}$.
Given the assignment $(i,j)\rightarrow\mathcal{K}$, $\mathcal{K}$
$\in\mathcal{O}_{\mathcal{B}}$ if $1\leq j\leq K_{i}$ and $\mathcal{K}$
$\in\mathcal{O}_{\mathcal{A}}$ otherwise. Observer $\mathcal{A}_{i}$ randomly
chooses input bit $x_{i}\in\{0,1\}$ and measures the observable $A_{x_{i}%
}^{(i)}$ on the qubit at hand. Observer $\mathcal{B}_{j}$ randomly chooses
input bit string $n_{B,j}$-bit string $y_{j}=y_{j,1}\ldots y_{j,n_{B,j}}$ and
measures the joint observable $B_{y_{j}}^{(j)}$ on the qubits at hand.
Hereafter the observer $\mathcal{K}$ performs the type-$X$ measurement if
$\mathcal{K}$ $\in\mathcal{O}_{X}$, where $X\in\{A,B\}$. To design the Bell
operators of Bell-type inequalities tailored for $\mathcal{N}_{Q}^{(n,M)}$, we
use the "cut-and-garfted" method introduced in \cite{hsu1}. Given the observer
$\mathcal{A}_{i}$ perfoming the type-$A$ measurement, assign the observable
$O_{A,l}^{(i)}\rightarrow\frac{(A_{0}^{(i)}+(-1)^{l_{i}}A_{1}^{(i)})}%
{2\omega_{l}^{(i)}}$, where $\omega_{l}^{(i)}=(\cos\theta_{i})^{\overline
{l_{i}}}(\sin\theta_{i})^{l_{i}}$ and $0<\theta_{i}<\frac{\pi}{2}$ for all
$i$. Given the observer $\mathcal{B}_{j}$ perfoming the type-$B$ measurement,
assign the joint observable $O_{B,l}^{(j)}\rightarrow B_{y_{j}(l)}^{(j)}$. As
a result, the associated Bell operators $\mathbf{B}_{l}$ from $g_{l}$ reads%
\begin{equation}
\mathbf{B}_{l}=%
%TCIMACRO{\dprod \nolimits_{j=1}^{M_{B}}}%
%BeginExpansion
{\displaystyle\prod\nolimits_{j=1}^{M_{B}}}
%EndExpansion%
%TCIMACRO{\dprod \nolimits_{i=1}^{M_{A}}}%
%BeginExpansion
{\displaystyle\prod\nolimits_{i=1}^{M_{A}}}
%EndExpansion
\frac{(A_{0}^{(i)}+(-1)^{l_{i}}A_{1}^{(i)})}{2}B_{y_{j}(l)}^{(j)}. \label{seg}%
\end{equation}
Notably, the assignment on $O_{A,l}^{(i)}$ for type-$A$ measurement and
construction of the associated Bell operators are different by the\ factor
$\omega_{l}^{(i)}$.

For example, in the two-source network $\mathcal{N}_{Q}^{(2,3)}$ with
parameters $M_{A}=2$, $M_{B}=1$ and $n_{B,1}=2$, let the quantum sources
$e_{1}^{(Q)}$and $e_{2}^{(Q)}$ in each emit the two-qubit Bell state and let
$\left\vert \psi_{1}\right\rangle =\left\vert \psi_{2}\right\rangle
=\left\vert Bell\right\rangle =$ $(\left\vert 00\right\rangle +\left\vert
11\right\rangle )/\sqrt{2}$, which is stabilized by the operators $ZZ$ and
$XX$. The qubits $(1,2)$ and $(2,2)$ are sent to $\mathcal{A}_{1}$ and
$\mathcal{A}_{2}$, respectively, while the qubits $(1,1)$ and $(2,1)$ are both
sent to $\mathcal{B}_{1}$. The useful local stabilizing operators
are$\ g_{l_{1}=0}^{(1)}=Z_{(1,1)}Z_{(1,2)}$, $g_{l_{1}=1}^{(1)}=X_{(1,1)}%
X_{(1,2)}$, $g_{l_{2}=0}^{(2)}=Z_{(2,1)}Z_{(2,2)}$, and $g_{l_{2}=1}%
^{(2)}=X_{(2,1)}X_{(2,2)}$. There are four useful blobal stabilzing operators:
$g_{ij}=g_{l_{1}=i}^{(1)}g_{l_{2}=j}^{(2)}=O_{A,ij}^{(1)}$ $O_{A,ij}%
^{(2)}O_{B,ij}^{(1)}$, $i$, $j=0$, $1$. For qubit $(1,2)$, assign
$O_{A,00}^{(1)}=O_{A,01}^{(1)}=Z_{(1,2)}\rightarrow\frac{(A_{0}^{(1)}%
+A_{1}^{(1)})}{2\cos\theta_{1}}$ and $O_{A,10}^{(1)}=O_{A,11}^{(1)}%
=X_{(1,2)}\rightarrow\frac{(A_{0}^{(1)}-A_{1}^{(1)})}{2\sin\theta_{1}}$. For
qubit $(2,2)$, assign $O_{A,00}^{(2)}=O_{A,10}^{(2)}=Z_{(2,2)}\rightarrow
\frac{(A_{0}^{(2)}+A_{1}^{(2)})}{2\cos\theta_{2}}$ and $O_{A,01}%
^{(2)}=O_{A,11}^{(2)}=X_{(2,2)}\rightarrow\frac{(A_{0}^{(2)}-A_{1}^{(2)}%
)}{2\sin\theta_{2}}$, where $0<\theta_{1},\theta_{2}<\frac{\pi}{2}$. For
qubits $(1,1)$ and $(2,1),$ let $y_{1}(l)=l$. In this case, assign
$O_{B,00}^{(1)}=Z_{(1,1)}Z_{(2,1)}\rightarrow B_{00}^{(1)}$, $O_{B,01}%
^{(1)}=Z_{(1,1)}X_{(2,1)}\rightarrow B_{01}^{(1)}$, $O_{B,10}^{(1)}%
=X_{(1,1)}Z_{(2,1)}\rightarrow B_{10}^{(1)}$ and $O_{B,11}^{(1)}%
=X_{(1,1)}X_{(2,1)}\rightarrow B_{11}^{(1)}$. To construct the associated Bell
operators $\mathbf{B}_{l}$ directly from $g_{l}$, the operators $Z_{(i,2)}$
and $X_{(i,2)}$ in $g_{l}$ are just replaced by $\frac{(A_{0}^{(i)}%
+A_{1}^{(i)})}{2}$ and $\frac{(A_{0}^{(i)}-A_{1}^{(i)})}{2}$, respectively,
and the observable $O_{B,l}^{(j)}$ is replaced by $B_{y_{j}(l)}^{(j)}$. As a
result, we have
\[
\mathbf{B}_{l}=(\frac{A_{0}^{(1)}+(-1)^{l_{1}}A_{1}^{(1)}}{2})(\frac
{A_{0}^{(2)}+(-1)^{l_{2}}A_{1}^{(2)}}{2})B_{l}^{(1)}.
\]

Next we explore the star-shaped network $\mathcal{N}_{Q}^{(n,n+1)}$ with
$n=M_{A}$, $M_{B}=1$ as a generalization of $\mathcal{N}_{Q}^{(2,3)}$. In this
configuration, the $i$-th quantum source emits the quantum state $\left\vert
\psi_{i}\right\rangle =\left\vert Bell\right\rangle $, and qubits $(i,2)$ and
$(i,1)$ are sent to $\mathcal{A}_{i}$ and $\mathcal{B}_{1}$, respectively, for
all $i=1,2,...,n$. Considering the $n$-bit string $l=l_{1},l_{2},...,l_{n}$
and defining $y_{1}(l)=l$, the global stabilizing operator is given by $g_{l}=%
%TCIMACRO{\dprod \nolimits_{i=1}^{n}}%
%BeginExpansion
{\displaystyle\prod\nolimits_{i=1}^{n}}
%EndExpansion
(Z_{(i,1)}Z_{(i,2)})^{\overline{l_{i}}}(X_{(i,1)}X_{(i,2)})^{l_{i}}%
=O_{B,l}^{(1)}%
%TCIMACRO{\dprod \nolimits_{i=1}^{n}}%
%BeginExpansion
{\displaystyle\prod\nolimits_{i=1}^{n}}
%EndExpansion
O_{A,l}^{(i)}$, where $O_{A,l}^{(i)}=Z_{(i,2)}^{\overline{l_{i}}}%
X_{(i,2)}^{l_{i}}$, $O_{B,l}^{(1)}=%
%TCIMACRO{\dprod \nolimits_{i=1}^{n}}%
%BeginExpansion
{\displaystyle\prod\nolimits_{i=1}^{n}}
%EndExpansion
Z_{(i,1)}^{\overline{l_{i}}}X_{(i,1)}^{l_{i}}$, and $\overline{l_{i}}%
=l_{i}+1\operatorname{mod}2$. By substituting $Z_{(i,1)}$ and $X_{(i,1)}$ in
$g_{l}$ with $\frac{(A_{0}^{(i)}+A_{1}^{(i)})}{2}$ and $\frac{(A_{0}%
^{(i)}-A_{1}^{(i)})}{2}$, respectively, and assigning $O_{B,l}^{(1)}%
\rightarrow B_{y_{1}(l)}^{(1)}$, we can derive the segmented Bell operator
$\mathbf{B}_{l}$ from $g_{l}$. Specifically, denote $\mathbf{B}_{i}^{(l_{i}%
)}=$ $B_{l_{i}}^{(i,1)}(\frac{A_{0}^{(i,2)}+(-1)^{l_{i}}A_{1}^{(i,2)}}{2})$,
where $B_{l_{i}}^{(i,1)}=Z_{(i,1)}^{\overline{l_{i}}}X_{(k,1)}^{l_{i}}$. By
grafting (tensoring) $n$ segmented Bell operaors $\mathbf{B}_{1}^{(l_{1}%
)},...,\mathbf{B}_{n}^{(l_{n})}$, we have $\mathbf{B}_{l}=%
%TCIMACRO{\dprod \nolimits_{i=1}^{n}}%
%BeginExpansion
{\displaystyle\prod\nolimits_{i=1}^{n}}
%EndExpansion
\mathbf{B}_{i}^{(l_{i})}$. Notably, in the $n=1$ case, the $\mathcal{N}%
_{Q}^{(0,1)}$ reduces to the standard Bell test scenario, where the CHSH-Bell
operator is the linear combination of $\mathbf{B}_{i}^{(0)}$ and
$\mathbf{B}_{i}^{(1)}$.

\textit{Bell inequalities }To construct Bell inequalities tailored to the
quantum product state $\left\vert \Psi\right\rangle $ with qubits distributed
in $\mathcal{N}_{Q}^{(n,M)}$, we explore the maximal local correlation in the
fully classical network $\mathcal{N}_{C}^{(n,M)}$ with the classical source
$e_{i}^{(C)}$ emitting the local hidden variable $\lambda_{i}$ for all
$i=1,...,n$. In the literature, the local hidden variables are distributed
among limited observers in $\mathcal{N}_{C}^{(n,M)}$. Here we relax the
constraint: considering any local hidden variable as a classical object, any
element in the set $\mathbf{\lambda}=\{\lambda_{1},\lambda_{2},...,\lambda
_{n}\}$ can be perfectly cloned and then spread throughout $\mathcal{N}%
_{C}^{(n,M)}$. That is, $\mathbf{\lambda}$ is accessible to any observers in
$\mathcal{N}_{C}^{(n,M)}$. Denote $a_{x_{i}}^{(i)}$ as the classical outcome
of the observable $A_{x_{i}}^{(i)}$ measured by $\mathcal{A}_{i}$. According
to the local realism therein, either $\left\vert (a_{x_{i}}^{(i)}+a_{x_{i}%
}^{(i)})/2\right\vert $ or $\left\vert (a_{x_{i}}^{(i)}-a_{x_{i}}%
^{(i)})/2\right\vert $ must be 1, and the other must be zero. That is,
$P(\left\vert \frac{a_{x_{i}}^{(i)}+a_{x_{i}}^{(i)}}{2}\right\vert
=1|\mathbf{\lambda})+P(\left\vert \frac{a_{x_{i}}^{(i)}-a_{x_{i}}^{(i)}}%
{2}\right\vert =1|\mathbf{\lambda})=1$ $\forall i$. In addition, given any
$\mathbf{\lambda}$, there must be one and only one $M_{A}$-bit binary string
$l$ that satisfies $%
%TCIMACRO{\dprod \nolimits_{i=1}^{M_{A}}}%
%BeginExpansion
{\displaystyle\prod\nolimits_{i=1}^{M_{A}}}
%EndExpansion
$ $\left\vert \frac{a_{0}^{(i)}(\mathbf{\lambda})+(-1)^{l_{i}}a_{1}%
^{(i)}(\mathbf{\lambda})}{2}\right\vert =1$, and $%
%TCIMACRO{\dprod \nolimits_{i=1}^{M_{A}}}%
%BeginExpansion
{\displaystyle\prod\nolimits_{i=1}^{M_{A}}}
%EndExpansion
$ $\left\vert \frac{a_{x_{i}}^{(i)}(\mathbf{\lambda})+(-1)^{l_{i}^{\prime}%
}a_{x_{i}}^{(i)}(\mathbf{\lambda})}{2}\right\vert =0$ for any other $M_{A}%
$-bit binary binary string $l^{\prime}\neq l$. Denote the probability
$P_{l}=P(%
%TCIMACRO{\dprod \nolimits_{i=1}^{M_{A}}}%
%BeginExpansion
{\displaystyle\prod\nolimits_{i=1}^{M_{A}}}
%EndExpansion
\left\vert \frac{a_{0}^{(i)}+(-1)^{l_{i}}a_{1}^{(i)}}{2}\right\vert
=1|\mathbf{\lambda})$, and we have%

\begin{equation}
P_{l}=\left\langle
%TCIMACRO{\dprod \nolimits_{i=1}^{M_{A}}}%
%BeginExpansion
{\displaystyle\prod\nolimits_{i=1}^{M_{A}}}
%EndExpansion
\left\vert \frac{a_{0}^{(i)}+(-1)^{l_{i}}a_{1}^{(i)}}{2}\right\vert
\right\rangle _{C},
\end{equation}
where $\left\langle \cdot\right\rangle $ denotes the average value of $\cdot$
. On the other hand, let the observer $\mathcal{B}_{j}$ in $\mathcal{N}%
_{C}^{(n,M)}$ measure the observable $B_{y_{j(l),k}}^{(j)}$ on the $k$-th
system at hand and then obtain the measurement outcome $b_{y_{j(l),k}}^{(j)}$
$\forall k=1,...,n_{B}^{(j)}$. Denote the joint observable $B_{y_{j(l)}}%
^{(j)}=%
%TCIMACRO{\dprod \nolimits_{k=1}^{n_{B,j}}}%
%BeginExpansion
{\displaystyle\prod\nolimits_{k=1}^{n_{B,j}}}
%EndExpansion
$ $B_{y_{j(l),k}}^{(j)}$ with the final outcome $b_{y_{j(l)}}^{(j)}=%
%TCIMACRO{\dprod \nolimits_{k=1}^{n_{B,j}}}%
%BeginExpansion
{\displaystyle\prod\nolimits_{k=1}^{n_{B,j}}}
%EndExpansion
$\ $b_{y_{j(l),k}}^{(j)}\in\{1,-1\}$. Consequently, the linear Bell inequality
reads
\begin{equation}%
%TCIMACRO{\dsum \nolimits_{l}}%
%BeginExpansion
{\displaystyle\sum\nolimits_{l}}
%EndExpansion
\left\langle \mathbf{B}_{l}\right\rangle _{C}\leq1. \label{linear}%
\end{equation}
The lines of the proof read
\begin{align}
&
%TCIMACRO{\dsum \nolimits_{l}}%
%BeginExpansion
{\displaystyle\sum\nolimits_{l}}
%EndExpansion
\left\langle \mathbf{B}_{l}\right\rangle _{C}\nonumber\\
&  =%
%TCIMACRO{\dsum \nolimits_{l}}%
%BeginExpansion
{\displaystyle\sum\nolimits_{l}}
%EndExpansion
\left\langle
%TCIMACRO{\dprod \nolimits_{i=1}^{M_{A}}}%
%BeginExpansion
{\displaystyle\prod\nolimits_{i=1}^{M_{A}}}
%EndExpansion
\frac{a_{0}^{(i)}+(-1)^{l_{i}}a_{1}^{(i)}}{2}%
%TCIMACRO{\dprod \nolimits_{j=1}^{M_{B}}}%
%BeginExpansion
{\displaystyle\prod\nolimits_{j=1}^{M_{B}}}
%EndExpansion
b_{y_{j}(l)}^{(j)}\right\rangle _{C}\nonumber\\
&  \leq%
%TCIMACRO{\dsum \nolimits_{l}}%
%BeginExpansion
{\displaystyle\sum\nolimits_{l}}
%EndExpansion
\left\vert \left\langle
%TCIMACRO{\dprod \nolimits_{i=1}^{M_{A}}}%
%BeginExpansion
{\displaystyle\prod\nolimits_{i=1}^{M_{A}}}
%EndExpansion
\frac{a_{0}^{(i)}+(-1)^{l_{i}}a_{1}^{(i)}}{2}%
%TCIMACRO{\dprod \nolimits_{j=1}^{M_{B}}}%
%BeginExpansion
{\displaystyle\prod\nolimits_{j=1}^{M_{B}}}
%EndExpansion
b_{y_{j}(l)}^{(j)}\right\rangle _{C}\right\vert \nonumber\\
&  \leq%
%TCIMACRO{\dsum \nolimits_{l}}%
%BeginExpansion
{\displaystyle\sum\nolimits_{l}}
%EndExpansion
\left\langle \left\vert
%TCIMACRO{\dprod \nolimits_{i=1}^{M_{A}}}%
%BeginExpansion
{\displaystyle\prod\nolimits_{i=1}^{M_{A}}}
%EndExpansion
\frac{a_{0}^{(i)}+(-1)^{l_{i}}a_{1}^{(i)}}{2}%
%TCIMACRO{\dprod \nolimits_{j=1}^{M_{B}}}%
%BeginExpansion
{\displaystyle\prod\nolimits_{j=1}^{M_{B}}}
%EndExpansion
b_{y_{j}(l)}^{(j)}\right\vert \right\rangle _{C}\nonumber\\
&  =%
%TCIMACRO{\dsum \nolimits_{l}}%
%BeginExpansion
{\displaystyle\sum\nolimits_{l}}
%EndExpansion
\left\langle \left\vert
%TCIMACRO{\dprod \nolimits_{i=1}^{M_{A}}}%
%BeginExpansion
{\displaystyle\prod\nolimits_{i=1}^{M_{A}}}
%EndExpansion
\frac{a_{0}^{(i)}+(-1)^{l_{i}}a_{1}^{(i)}}{2}\right\vert \right\rangle
_{C}\nonumber\\
&  =%
%TCIMACRO{\dsum \nolimits_{l}}%
%BeginExpansion
{\displaystyle\sum\nolimits_{l}}
%EndExpansion
P_{l}\nonumber\\
&  =1 \label{LBI}%
\end{align}
Here, the second inequality arises from the inequality $\left\vert
\left\langle \cdot\right\rangle \right\vert \leq$ $\left\langle \left\vert
\cdot\right\vert \right\rangle $. In addition, we can construct the non-linear
Bell inequality that reads
\begin{equation}%
%TCIMACRO{\dsum \nolimits_{l}}%
%BeginExpansion
{\displaystyle\sum\nolimits_{l}}
%EndExpansion
\left\langle \mathbf{B}_{l}\right\rangle _{C}^{p}\leq\mathbf{B}_{C}%
^{(M_{A},p)}, \label{BellInq}%
\end{equation}
where $\mathbf{B}_{C}^{(M_{A},p)}=2^{M_{A}(1-p)}$. The proof proceeds as
follows: since $%
%TCIMACRO{\dsum \nolimits_{l}}%
%BeginExpansion
{\displaystyle\sum\nolimits_{l}}
%EndExpansion
\left\langle \mathbf{B}_{l}\right\rangle _{C}^{p}\leq%
%TCIMACRO{\dsum \nolimits_{l}}%
%BeginExpansion
{\displaystyle\sum\nolimits_{l}}
%EndExpansion
\left\vert \left\langle \mathbf{B}_{l}\right\rangle _{C}\right\vert ^{p}$ and
\begin{align}
&
%TCIMACRO{\dsum \nolimits_{l}}%
%BeginExpansion
{\displaystyle\sum\nolimits_{l}}
%EndExpansion
\left\vert \left\langle \mathbf{B}_{l}\right\rangle _{C}\right\vert
^{p}\nonumber\\
&  =%
%TCIMACRO{\dsum \nolimits_{l}}%
%BeginExpansion
{\displaystyle\sum\nolimits_{l}}
%EndExpansion
\left\vert \left\langle
%TCIMACRO{\dprod \nolimits_{i=1}^{M_{A}}}%
%BeginExpansion
{\displaystyle\prod\nolimits_{i=1}^{M_{A}}}
%EndExpansion
\frac{a_{0}^{(i)}+(-1)^{l_{i}}a_{1}^{(i)}}{2}%
%TCIMACRO{\dprod \nolimits_{j=1}^{M_{B}}}%
%BeginExpansion
{\displaystyle\prod\nolimits_{j=1}^{M_{B}}}
%EndExpansion
b_{y_{j}(l)}^{(j)}\right\rangle _{C}\right\vert ^{p}\nonumber\\
&  \leq%
%TCIMACRO{\dsum \nolimits_{l}}%
%BeginExpansion
{\displaystyle\sum\nolimits_{l}}
%EndExpansion
\left\langle \left\vert
%TCIMACRO{\dprod \nolimits_{i=1}^{M_{A}}}%
%BeginExpansion
{\displaystyle\prod\nolimits_{i=1}^{M_{A}}}
%EndExpansion
\frac{a_{0}^{(i)}+(-1)^{l_{i}}a_{1}^{(i)}}{2}%
%TCIMACRO{\dprod \nolimits_{j=1}^{M_{B}}}%
%BeginExpansion
{\displaystyle\prod\nolimits_{j=1}^{M_{B}}}
%EndExpansion
b_{y_{j}(l)}^{(j)}\right\vert \right\rangle _{C}^{p}\nonumber\\
&  =%
%TCIMACRO{\dsum \nolimits_{l}}%
%BeginExpansion
{\displaystyle\sum\nolimits_{l}}
%EndExpansion
P_{l}^{p}\nonumber\\
&  \leq2^{M_{A}(1-p)}, \label{cc}%
\end{align}
where, using the Lagrange multiplier method with the constraint $%
%TCIMACRO{\dsum \nolimits_{l}}%
%BeginExpansion
{\displaystyle\sum\nolimits_{l}}
%EndExpansion
P_{l}=1$, it is easy to verify that the equality of the second inequality
holds if $P_{l}=2^{-M_{A}}$ for all $l.$ Note that the linear Bell inequality
with $p=1$ in (\ref{cc}) can be regarded as the stronger form of (\ref{linear}).

\textit{Quantum and no-signaling upper bounds} To determine the quantum upper
bound of the sum $%
%TCIMACRO{\dsum \nolimits_{l}}%
%BeginExpansion
{\displaystyle\sum\nolimits_{l}}
%EndExpansion
\left\langle \mathbf{B}_{l}\right\rangle _{Q}^{p}$, we employ the
sum-of-squares decomposition \cite{sos}. We define the operator $M_{l}$ such
that \ $M_{l}$\ $\left\vert \Phi\right\rangle =$ $(%
%TCIMACRO{\dprod \nolimits_{i=1}^{M_{A}}}%
%BeginExpansion
{\displaystyle\prod\nolimits_{i=1}^{M_{A}}}
%EndExpansion
O_{A,l}^{(i)}\left\vert \Phi\right\rangle )^{p}-(%
%TCIMACRO{\dprod \nolimits_{j=1}^{M_{B}}}%
%BeginExpansion
{\displaystyle\prod\nolimits_{j=1}^{M_{B}}}
%EndExpansion
O_{B,l}^{(j)}\left\vert \Phi\right\rangle )^{p}$ \cite{09,10,011}. Assign the
observables $O_{A,l}^{(i)}=Z_{(i,2)}^{\overline{l_{i}}}X_{(i,2)}^{l_{i}%
}\rightarrow\frac{(A_{0}^{(i)}+(-1)^{l_{i}}A_{1}^{(i)})}{2\omega_{l}^{(i)}}$
and $O_{B,l}^{(j)}\rightarrow B_{y_{j}(l)}^{(j)}$, where $\omega_{l}%
^{(i)}=(\cos\theta_{i})^{\overline{l_{i}}}(\sin\theta_{i})^{l_{i}}$ and
$0<\theta_{i}<\frac{\pi}{2}$ for all $i$. Denote the semi-definite operator
$\gamma\mathbf{=}%
%TCIMACRO{\dsum \nolimits_{l}}%
%BeginExpansion
{\displaystyle\sum\nolimits_{l}}
%EndExpansion
$\ $\omega_{l}^{p}M_{l}^{\dagger}M_{l}$ with $\omega_{l}=$ $%
%TCIMACRO{\dprod \nolimits_{i=1}^{M_{A}}}%
%BeginExpansion
{\displaystyle\prod\nolimits_{i=1}^{M_{A}}}
%EndExpansion
\omega_{l}^{(i)}>0$. Given that $\left\langle \omega_{l}^{p}M_{l}^{\dagger
}M_{l}\right\rangle \geq0$ and $O_{A,l}^{(i)\dagger}O_{A,l}^{(i)}%
=O_{B,l}^{(j)\dagger}O_{B,l}^{(j)}=1$, we have $\omega_{l}^{p}\geq\left\langle
\Phi\left\vert \ \mathbf{B}_{l}\right\vert \Phi\right\rangle ^{p}$and hence $%
%TCIMACRO{\dsum \nolimits_{_{l}}}%
%BeginExpansion
{\displaystyle\sum\nolimits_{_{l}}}
%EndExpansion
\omega_{l}^{p}\geq%
%TCIMACRO{\dsum \nolimits_{_{l}}}%
%BeginExpansion
{\displaystyle\sum\nolimits_{_{l}}}
%EndExpansion
\left\langle \Phi\left\vert \ \mathbf{B}_{l}\right\vert \Phi\right\rangle
^{p}$ for any quantum state $\left\vert \Phi\right\rangle $. On the other
hand, since $M_{l}\left\vert \Psi\right\rangle =0$ and $\left\langle
\Psi\left\vert M_{l}^{\dagger}M_{l}\right\vert \Psi\right\rangle =0$, it
follows that $\omega_{l}^{p}$ $=\left\langle \Psi\left\vert M_{l}^{\dagger
}M_{l}\right\vert \Psi\right\rangle ^{p}=\max_{\left\vert \Phi\right\rangle
}\left\langle \Phi\left\vert \ \mathbf{B}_{l}\right\vert \Phi\right\rangle
^{p}$. Denote the quantum upper bound of the Bell inequality as $\mathbf{B}%
_{Q}^{(M_{A},p)}$, the proof of $\mathbf{B}_{Q}^{(M_{A},p)}=2^{M_{A}%
(1-\frac{p}{2})}$ proceeds as follows:%
\begin{align}
&  \max_{\left\vert \Phi\right\rangle }%
%TCIMACRO{\dsum \nolimits_{_{l}}}%
%BeginExpansion
{\displaystyle\sum\nolimits_{_{l}}}
%EndExpansion
\left\langle \mathbf{B}_{l}\right\rangle _{\Phi}^{p}\nonumber\\
&  =%
%TCIMACRO{\dsum \nolimits_{_{l}}}%
%BeginExpansion
{\displaystyle\sum\nolimits_{_{l}}}
%EndExpansion
\left\langle \mathbf{B}_{l}\right\rangle _{\Psi}^{p}\nonumber\\
&  =%
%TCIMACRO{\dsum \nolimits_{_{l}}}%
%BeginExpansion
{\displaystyle\sum\nolimits_{_{l}}}
%EndExpansion
\omega_{l}^{p}\nonumber\\
&  =%
%TCIMACRO{\dprod \nolimits_{i=1}^{M_{A}}}%
%BeginExpansion
{\displaystyle\prod\nolimits_{i=1}^{M_{A}}}
%EndExpansion
(\cos^{p}\theta_{i}+\sin^{p}\theta_{i})\nonumber\\
&  \leq2^{M_{A}(1-\frac{p}{2})}. \label{maxxx}%
\end{align}
Note that the equality in the inequality $(\cos^{p}\theta+\sin^{p}\theta
)\leq2^{1-\frac{p}{2}}$ holds if $\theta=\frac{\pi}{4}$. That is,
$\max_{\left\vert \Phi\right\rangle }\left\langle \Phi\left\vert
\ \mathbf{B}_{l}\right\vert \Phi\right\rangle ^{p}=2^{-M_{A}\frac{p}{2}}.$ In
the standard Bell test with $n=M_{A}=M_{B}=1$, denote the two segmented Bell
operators $\mathbf{B}_{l=0}=\frac{B(A+A^{\prime})}{2}$, $\mathbf{B}%
_{l=1}=\frac{B^{\prime}(A-A^{\prime})}{2}$ and $p=2$, and then we reach the
quadratic inequality in the quantum region $\left\langle \frac{B(A+A^{\prime
})}{2}\right\rangle _{Q}^{2}+\left\langle \frac{B^{\prime}(A-A^{\prime})}%
{2}\right\rangle _{Q}^{2}\leq1$, which is different from that in
\cite{pan,pan1}.

For the no-signaling systems in $\mathcal{N}_{NS}^{(n,M)}$, it is presumed
that the algebraic maximum of any segmented Bell operators can be reached,
meaning $\left\langle \mathbf{B}_{l}\right\rangle _{NS}=1$ for all $l$. For
instance, with a single source emitting a Popescu-Rohrlich (PR) box comprising
two non-signaling systems, we have $\left\langle \mathbf{B}_{1}^{(l=0)}%
\right\rangle _{NS}=\left\langle \frac{B_{0}^{(1)}(A_{0}^{(1)}+A_{1}^{(1)}%
)}{2}\right\rangle _{NS}=1$ and $\left\langle \mathbf{B}_{1}^{(l=1)}%
\right\rangle _{NS}=\left\langle \frac{B_{1}^{(1)}(A_{0}^{(1)}-A_{1}^{(1)}%
)}{2}\right\rangle _{NS}=1$ \cite{66}. In a star-shaped network $\mathcal{N}%
_{NS}^{(n,n+1)}$ with each non-signaling source emitting a PR box, we have
$\left\langle \mathbf{B}_{l}\right\rangle _{NS}=\left\langle
%TCIMACRO{\dprod \nolimits_{i=1}^{n}}%
%BeginExpansion
{\displaystyle\prod\nolimits_{i=1}^{n}}
%EndExpansion
\mathbf{B}_{i}^{(l_{i})}\right\rangle _{NS}=1.$ As a result, the algebraic
maximum of the nonlinear Bell inequality (\ref{BellInq}), denoted by
$\mathbf{B}_{NS}^{(M_{A},p)}$, can be achieved in $\mathcal{N}_{NS}^{(n,M)}$,
yielding:%
\begin{equation}
\mathbf{B}_{NS}^{(M_{A},p)}=%
%TCIMACRO{\dsum \nolimits_{_{l}}}%
%BeginExpansion
{\displaystyle\sum\nolimits_{_{l}}}
%EndExpansion
\left\langle \mathbf{B}_{l}\right\rangle _{NS}^{p}=2^{M_{A}}. \label{algem}%
\end{equation}
Some remarks are in order. Notably, the critical values $\mathbf{B}%
_{C}^{(M_{A},p)}$, $\mathbf{B}_{Q}^{(M_{A},p)}$ and $\mathbf{B}_{NS}%
^{(M_{A},p)}$ in the nonlinear Bell inequalities increase expotentially with
$M_{A}$ and are independent of $n$\ and $M_{B}$. This highlights the
significant role of the number of type-$A$ measurements. Secondly, since the
type-$B$ measurements play a minor role in evaluating network nonlocality,
without loss of generality, it can be presumed that $\mathcal{B}_{1}%
=\cdots=\mathcal{B}_{M_{B}}=\mathcal{B}$ who measures the joint observable $%
%TCIMACRO{\dprod \nolimits_{j=1}^{M_{B}}}%
%BeginExpansion
{\displaystyle\prod\nolimits_{j=1}^{M_{B}}}
%EndExpansion
B_{y_{j}(l)}^{(j)}$ \cite{star1,star2}. Consequently, the achievable
correlation strength in $\mathcal{N}^{(n,M)}$ with arbitrary topology can be
modeled as a star-shaped network. Thirdly, in the examples discussed, the
quantum sources emit two-qubit Bell states. In appendix, we explore the
achievable correlations in $\mathcal{N}_{Q}^{(n,M)}$ with sources emitting
multi-qubit Greenberger-Horne-Zeilinger (GHZ) states. As a result, when
quantum sources $\mathcal{N}_{Q}^{(n,M)}$ are restricted to emitting
stabilizer states, any of the upper bounds $\mathbf{B}_{C}^{(M_{A},p)}$,
$\mathbf{B}_{Q}^{(M_{A},p)}$ and $\mathbf{B}_{NS}^{(M_{A},p)}$ is still
independent of prepared stabilizer states and depend only on $M_{A}$, since
$2^{M_{A}}$ segmented Bell operators are\ involved in the proposed Bell
inequalities. For instance, in the previous examples on the networks
$\mathcal{N}_{C}^{(2,3)}$ and $\mathcal{N}_{Q}^{(2,3)}$, there are four
segmented Bell operators. By setting $p=\frac{1}{2}$, the proposed Bell
inequality reads $%
%TCIMACRO{\dsum \nolimits_{l}}%
%BeginExpansion
{\displaystyle\sum\nolimits_{l}}
%EndExpansion
\left\langle \mathbf{B}_{l}\right\rangle _{C}^{1/2}\leq2$. On the other hand,
regarding the bilocality inequality that reads $%
%TCIMACRO{\dsum \nolimits_{_{l=00,11}}}%
%BeginExpansion
{\displaystyle\sum\nolimits_{_{l=00,11}}}
%EndExpansion
\left\langle \mathbf{B}_{l}\right\rangle _{C}^{1/2}\leq1$ \cite{5-2,bii},
which includes only two segmented Bell operators, the classical upper bound
can be achieved by letting either $P_{00}=1$ or $P_{11}=1$. In this case, the
quantum upper bound can be reached by preparing either the pure state
$\left\vert Bell\right\rangle ^{\otimes2}$ or the 4-qubit Smolin mixed state,
which is bound-entangled and bi-separable \cite{Smolin}. As\ for the proposed
Bell inequality \ $%
%TCIMACRO{\dsum \nolimits_{_{l}}}%
%BeginExpansion
{\displaystyle\sum\nolimits_{_{l}}}
%EndExpansion
\left\langle \mathbf{B}_{l}\right\rangle _{C}^{1/2}\leq\mathbf{B}%
_{C}^{(2,1/2)}=1$, the quantum upper bound can achieved achieved only by the
pure state $\left\vert Bell\right\rangle ^{\otimes2}$ \cite{hsu2,acin1}.

\textit{Star-shaped hybrid networks. }In the star-shaped hybrid network
$\mathcal{N}_{H}^{\left\langle u,v,w\right\rangle }$, let $e_{i}^{(C)}$ be a
classical source that emits $N_{i}^{(C)}$classical systems with the local
hidden variable $\lambda_{i}$ for all $i=1,...,u$; let $e_{u+j}^{(Q)}$ be a
quantum source that $N_{j}^{(Q)}$-qubits qauntum state $\left\vert \psi
_{j}\right\rangle $ for all $j=1,...v$; let $e_{u+v+k}^{(NS)}$ be a
\ no-signaling source that emits $N_{u+v+k}^{(NS)}$ no-signaling systems for
all $k=1,...,w=n-u-v$. In the following, let $N_{i}^{(C)}=N_{j}^{(Q)}%
=N_{k}^{(NS)}=2$ and $n=M_{A}$ The observers $\mathcal{A}_{1}$, $\mathcal{A}%
_{2}$,..., $\mathcal{A}_{u}$ and $\mathcal{B}_{1}$\ each receive the set of
the local hidden variables $\mathbf{\lambda=\{}\lambda_{i}$,..., $\lambda
_{u}\}$; $\mathcal{A}_{u+j}$ and $\mathcal{B}_{1}$ each receive a qubit from
$e_{u+j}^{(Q)}$ ; and $\mathcal{A}_{u+v+k}$ and $\mathcal{B}_{1}$ each receive
a no-signaling particle from $e_{u+v+k}^{(NS)}$. Each observer $\mathcal{A}%
_{i}$ measures the observable $A_{x_{i}}^{(i)}$ on the accessible system with
particle index $(i,2)$. Associated with the $M_{A}$-bit string $l$, Bob
measures the observable $B_{l}=B_{l}^{(C)}B_{l}^{(Q)}B_{l}^{(NS)}$ , where
$B_{l}^{(C)}$, $B_{l}^{(Q)}$, and $B_{l}^{(NS)}$ are the multi-particle
observables on the classical, quantum, no-signaling systems on the particles
with index sets $\{(1,1)$, ...,$(u,1)\}$, $\{(u+1,1)$, ...,$(u+v,1)\}$, and
$\{(u+v+1,1)$, ...,$(n,1)\}$, respectively. Here $l$ can be decomposed as
$l_{C}||l_{Q}||l_{NS}$, where $l_{C}$, $l_{Q}$ and $l_{NS}$ are $u$-bit,
$v$-bit, and $w$-bit strings, respectively. Denote the segmented Bell
operators for the associated bit string $l$ as $\mathbf{B}_{l}^{(u,v,w)}%
=\xi_{l}^{(C)}\xi_{l}^{(Q)}\xi_{l}^{(NS)}$, where
\begin{equation}
\xi_{l}^{(C)}=B_{l_{C}}^{(C)}%
%TCIMACRO{\dprod \nolimits_{i=1}^{u}}%
%BeginExpansion
{\displaystyle\prod\nolimits_{i=1}^{u}}
%EndExpansion
\frac{(A_{0}^{(i)}+(-1)^{l_{i}}A_{1}^{(i)})}{2}, \label{f1}%
\end{equation}%
\begin{equation}
\xi_{l}^{(Q)}=B_{l_{Q}}^{(Q)}%
%TCIMACRO{\dprod \nolimits_{j=u+1}^{u+v}}%
%BeginExpansion
{\displaystyle\prod\nolimits_{j=u+1}^{u+v}}
%EndExpansion
\frac{(A_{0}^{(j)}+(-1)^{l_{j}}A_{1}^{(j)})}{2}, \label{f2}%
\end{equation}
and
\begin{equation}
\xi_{l}^{(NS)}=B_{l_{NS}}^{(NS)}%
%TCIMACRO{\dprod \nolimits_{j=u+v+1}^{n}}%
%BeginExpansion
{\displaystyle\prod\nolimits_{j=u+v+1}^{n}}
%EndExpansion
\frac{(A_{0}^{(k)}+(-1)^{l_{k}}A_{1}^{(k)})}{2}. \label{f3}%
\end{equation}
As a result,
\begin{equation}
\left\langle \mathbf{B}_{l}^{(u,v,w)}\right\rangle =\left\langle \xi_{l}%
^{(C)}\right\rangle _{C}\left\langle \xi_{l}^{(Q)}\right\rangle _{Q}%
\left\langle \xi_{l}^{(NS)}\right\rangle _{NS}, \label{final}%
\end{equation}
which indicates that the correlations in $\mathcal{N}_{H}^{\left\langle
u,v,w\right\rangle }$ can be decomposed as the product of the correlations of
three star-shaped networks $\mathcal{N}_{C}^{(u,u+1)}$, \bigskip
$\mathcal{N}_{Q}^{(v,v+1)}$ and $\mathcal{N}_{NS}^{(w,w+1)}$. Before
proceeding further, we introduce the following inequality%

\begin{equation}
\forall z_{\left[  l\right]  }^{(X)}\geq0;\text{ }%
%TCIMACRO{\dsum \nolimits_{\left[  l\right]  =0}^{2^{M_{A}-1}}}%
%BeginExpansion
{\displaystyle\sum\nolimits_{\left[  l\right]  =0}^{2^{M_{A}-1}}}
%EndExpansion
(z_{\left[  l\right]  }^{(C)}z_{\left[  l\right]  }^{(Q)}z_{\left[  l\right]
}^{(NS)})^{\frac{1}{\alpha}}\leq%
%TCIMACRO{\dprod \nolimits_{X=C,Q,NS}}%
%BeginExpansion
{\displaystyle\prod\nolimits_{X=C,Q,NS}}
%EndExpansion
(%
%TCIMACRO{\dsum \nolimits_{\left[  l\right]  =0}^{2^{M_{A}-1}}}%
%BeginExpansion
{\displaystyle\sum\nolimits_{\left[  l\right]  =0}^{2^{M_{A}-1}}}
%EndExpansion
z_{\left[  l\right]  }^{(X)})^{\frac{1}{\alpha}}, \label{tt}%
\end{equation}
where $\left[  l\right]  $ denotes the binary number of the bit-string $l$.
Notably, the equality in (\ref{tt}) holds if $z_{0}^{(X)}=$ $z_{2}%
^{(X)}=\cdots=$ $z_{2^{M_{A}}-1}^{(X)}$ for all $X\in\{C,Q,NS\}$ \cite{ineq}.
Let $z_{\left[  l\right]  }^{(X)}$ in (\ref{tt}) be $\left\langle \xi
_{l}^{(X)}\right\rangle _{X}^{\alpha p}$, where (i) $\alpha=3$ if $uvw\neq0$;
(ii) $\alpha=2$ if one of $u$, $v$, or $w$ is 0 and the others are nonzero,
and (iii) $\alpha=1$ if two of $u$, $v$, or $w$ are $0$ and hence the network
$\mathcal{N}_{H}^{(u,v,w)}$ is reduced to one of these three networks
$\mathcal{N}_{C}^{(n,M_{A}+1)}$, $\mathcal{N}_{Q}^{(n,M_{A}+1)}$, or
$\mathcal{N}_{NS\text{ }}^{(n,M_{A}+1)}$. Denote the maximum correlation
strength in $\mathcal{N}_{H}^{\left\langle u,v,w\right\rangle }$ as
$\mathbf{B}_{\max}^{(u,v,w)}=\max%
%TCIMACRO{\dsum \nolimits_{l}}%
%BeginExpansion
{\displaystyle\sum\nolimits_{l}}
%EndExpansion
\left\langle \mathbf{B}_{l}^{(u,v,w)}\right\rangle ^{p}$ and, according to
(\ref{cc}-\ref{algem}), we have
\begin{equation}
\mathbf{B}_{\max}^{(u,v,w)}=2^{M_{A}-p(u+\frac{1}{2}v)}\text{,} \label{result}%
\end{equation}
To reach $\mathbf{B}_{\max}^{(u,v,w)}$, we have the expectation value
$\left\langle \mathbf{B}_{l}^{(u,v,w)}\right\rangle =2^{-p(u+\frac{v}{2})}$
with setting $\left\langle \xi_{l}^{(C)}\right\rangle _{C}=2^{-u}$,
$\left\langle \xi_{l}^{(Q)}\right\rangle _{Q}=2^{-\frac{v}{2}}$, and
$\left\langle \xi_{l}^{(NS)}\right\rangle _{NS}=1$ in (\ref{final}) for all
$l$. It is noteworthy that
\begin{equation}
\mathbf{B}_{\max}^{(u,n-u,0)}\leq\mathbf{B}_{\max}^{(u,v,w)}\leq
\mathbf{B}_{\max}^{(u,0,n-u)}, \label{r1}%
\end{equation}
which indicates that, regarding the number of local sources $u$ as a constant,
the maximum correlation strength of $\mathcal{N}_{H}^{(u,v,w)}$ can be
achieved by letting the other $n-u$ nonlocal sources emit PR boxes.

Finally, we state the main result of this work.

\textit{Under the constraint} $u+v+w=u^{\prime}+v^{\prime}+w^{\prime}=M_{A}%
$\textit{, the achievable correlations in }$\mathcal{N}_{H}^{(u,v,w)}$
\textit{cannot be reproduced by that in }$\mathcal{N}_{H}^{(u^{\prime
},v^{\prime},w^{\prime})}$ \textit{iff \ }$\mathbf{B}_{\max}^{(u,v,w)}%
>\mathbf{B}_{\max}^{(u^{\prime},v^{\prime},w^{\prime})}$ and hence
\begin{equation}
u+\frac{1}{2}v<u^{\prime}+\frac{1}{2}v^{\prime}. \label{main}%
\end{equation}
Notably, considering two-source networks with $M_{A}=2$, the correlation in
the full-quntum network $\mathcal{N}_{H}^{\left\langle 0,2,0\right\rangle
}=\mathcal{N}_{Q}^{(2,3)}$ \textit{can} be generated using the hybrid network
$\mathcal{N}_{H}^{\left\langle 1,0,1\right\rangle }$ since inequality
(\ref{main}) is violated by setting $u=w=v^{\prime}=0$, $u^{\prime}=w^{\prime
}=1$, and $v=2$. The achievable correlations in the full-qauntum network
$\mathcal{N}_{H}^{\left\langle 0,M_{A},0\right\rangle }=\mathcal{N}%
_{Q}^{(n,M_{A}+1)}$ cannot be reproduced in $\mathcal{N}_{H}^{\left\langle
u^{\prime},0,w^{\prime}\right\rangle }$ if $\frac{1}{2}M_{A}<u^{\prime}$.

Recently, the concept of full network nonlocality has been
proposed\ \cite{fnn}. A network is fully nonlocal if and only if the
achievable correlation strength cannot be modeled by allowing at least one
local source, while all other sources are general nonlocal ones. Here we
generalize full network nonlocality as follows. The network $\mathcal{N}%
_{H}^{\left\langle u,v,w\right\rangle }$ is called\textit{ }$t$-nonlocal if
and only if it cannot be modeled with at least\textit{ }$t$ classical sources,
while the other ($n-t$) sources are nonlocal. Specifically, the hybrid network
$\mathcal{N}_{H}^{\left\langle u,v,w\right\rangle }$ is $t$-nonlocal if and
only if $\max%
%TCIMACRO{\dsum \nolimits_{l}}%
%BeginExpansion
{\displaystyle\sum\nolimits_{l}}
%EndExpansion
\left\langle \mathbf{B}_{l}^{(u,v,w)}\right\rangle ^{p}>\mathbf{B}_{\max
}^{(t,0,n-t)}$. According to (\ref{main}), the achievable correlation strength
in fully quantum network $\mathcal{N}_{H}^{\left\langle 0,n,0\right\rangle
}=\mathcal{N}_{Q}^{(n,n+1)}$cannot be model by $\mathcal{N}_{H}^{\left\langle
t,0,M_{A}-t\right\rangle }$ if $2t>n$. As a result, the two-source networks
never be fully nonlocal \cite{fnn}.

In conclusion, we systematically study the maximal correlation strength of a
generic hybrid network. Therein, we study the associated fully-quantum network
and propose the Bell inequalities tailored to the prepared stabilizer states.
As shown in (\ref{result}), the maximal correlations of hybrid networks are
determined by the number of the type-$A$ measurements, $M_{A}$, and the
numbers of classical and quantum sources, $u$ and $v$. Notably, there are two
constraints on the type-$A$ measurements. One is listed in (\ref{criteria}),
and the other is that the type-$A$ measurement is performed on a single qubit.
Consequently, not all stabilizing operators are useful for designing the
associate Bell operators. In the future study, we aim to explore Bell
nonlocality in networks without these constraints. For example, if
$O_{A,l}^{(i)}\in\{I,$ $X,$ $Y,$ $Z\}$, the elegant joint measurement with
four measurement settings may be exploited for the type-$A$ measurements
\cite{ejm1, ejm2,ejm3}. In addition, we consider the multi-qubit type-$A$
measurement in the fully-quantum networks in appendix. We plan to investigate
variant correlations in the hybrid networks, where the correlation strengths
may be state-dependent.

\section{Acknowledgement}

The work is financially supported by National Science and Technology Council
(NSTC) with Grant No. NSTC 113-2112-M-033 -006.

\section{Appendix}

In the examples of the main text, the quantum sources are presumed to emit the
two-qubit Bell states. We extend our analysis to the scenario where the
quantum sources allow for emitting the multi-qubit stabilizer states. Let the
quantum source of the star-shaped network $\mathcal{N}_{Q}^{(n,n+1)}$ each
emit the maximally entangled state $\left\vert Bell\right\rangle $. Without
loss of generality, we study the star-shaped network $\mathcal{N}%
_{Q}^{(n,n+2)}$ with the quantum sources $e_{n}^{(Q)}$ in the star-shaped
network $\mathcal{N}_{Q}^{(n,n+1)}$ emitting the three-qubit GHZ state
$\left\vert GHZ_{3}\right\rangle =\frac{1}{\sqrt{2}}(\left\vert
000\right\rangle -\left\vert 111\right\rangle )$ instead of $\left\vert
Bell\right\rangle $. The qubits $(n,1)$ and $(n,2)$ are still sent to
$\mathcal{B}_{1}$, $\mathcal{A}_{n}$, respectively, and the qubit $(n,3)$ is
sent to$\ \mathcal{A}_{n+1}$. In this case, $M_{A}=n+1$ and $M_{B}=n+1$, and
$l$ denote ($n+1$)-bit string $l_{1}\cdots l_{n}l_{n+1}$. The stabilizer state
$\left\vert GHZ_{3}\right\rangle $ can be stabilized by $XYY$, $YXY$, $YYX$,
and $-XXX$. The $n-1$ segmented Bell operators $\mathbf{B}_{1}^{(l_{1})}$,
$...$, $\mathbf{B}_{n}^{(l_{n-1})}$ are the same as those tailored to the
star-shaped network $\mathcal{N}_{Q}^{(n,n+1)}$. In addition, assign the
measurement observables $Y_{(n,1)}^{l_{n}\oplus l_{n+1}}X_{(n,1)}%
^{\overline{l_{n}\oplus l_{n+1}}}\rightarrow B_{l_{n}l_{n+1}}^{(n,1)}%
=B_{l_{n}\oplus l_{n+1}}^{(n,1)}$, $Y_{(n,j)}\rightarrow\frac{(A_{0}%
^{(n+j-2)}+A_{1}^{(n+j-2)})}{2\cos\theta_{n+j-2}}$ and $X_{(n,j)}%
\rightarrow\frac{(A_{0}^{(n+j-2)}-A_{1}^{(n+j-2)})}{2\sin\theta_{n+j-2}}$,
where $j=2$, $3$ and $\oplus$ denotes the modulo-2 addition. Denote the
operator $\mathbf{B}_{n}^{(l_{n}l_{n+1})}=(-1)^{l_{n}\cdot l_{n+1}}%
B_{l_{n}l_{n+1}}^{(n,1)}\frac{(A_{0}^{(n)}+(-1)^{l_{n}}A_{1}^{(n)})}{2}%
\frac{(A_{0}^{(n+1)}+(-1)^{l_{n+1}}A_{1}^{(n+1)})}{2}$, and the three-qubit
Mermin inequalitiy reads $\left\langle
%TCIMACRO{\dsum \nolimits_{l_{n},l_{n+1}=0}^{1}}%
%BeginExpansion
{\displaystyle\sum\nolimits_{l_{n},l_{n+1}=0}^{1}}
%EndExpansion
\mathbf{B}_{n}^{(l_{n}l_{n+1})}\right\rangle \overset{LHV}{\leq}1$
\cite{Mermin}. \ As a result, the revised segmented Bell operator reads
$\mathbf{B}_{l}=$ $(%
%TCIMACRO{\dprod \nolimits_{i=1}^{n-1}}%
%BeginExpansion
{\displaystyle\prod\nolimits_{i=1}^{n-1}}
%EndExpansion
\mathbf{B}_{i}^{(l_{i})})\mathbf{B}_{n}^{(l_{n}l_{n+1})}$ with the associated
stabilizing operator $g_{l}=(-1)^{l_{n}\cdot l_{n+1}}$ $(%
%TCIMACRO{\dprod \nolimits_{i=1}^{n-1}}%
%BeginExpansion
{\displaystyle\prod\nolimits_{i=1}^{n-1}}
%EndExpansion
(Z_{(i,1)}Z_{(i,2)})^{\overline{l_{i}}}(X_{(i,1)}X_{(i,2)})^{l_{i}}%
)Y_{(n,1)}^{l_{n}\oplus l_{n+1}}X_{(n,1)}^{\overline{l_{n}\oplus l_{n+1}}%
}Y_{(n,2)}^{\overline{l_{n}}}X_{(n,2)}^{l_{n}}Y_{(n,3)}^{\overline{l_{n+1}}%
}X_{(n,3)}^{l_{n+1}}$. Similarly, the quantum upper bound $\mathbf{B}%
_{Q}^{(M_{A},p)}$ of the Bell inequality $%
%TCIMACRO{\dsum \nolimits_{l}}%
%BeginExpansion
{\displaystyle\sum\nolimits_{l}}
%EndExpansion
\left\vert \left\langle \mathbf{B}_{l}\right\rangle \right\vert ^{p}$ tailored
to the state $\left\vert Bell\right\rangle ^{\otimes n-1}\left\vert
GHZ_{3}\right\rangle $ distributed in $\mathcal{N}_{Q}^{(n,n+2)}$ is also
$2^{M_{A}(1-\frac{p}{2})}$ that can be achieved by setting $\theta_{n}%
=\theta_{n+1}=\frac{\pi}{4}$. It is easy to verify that the $\mathbf{B}%
_{C}^{(M_{A},p)}=2^{M_{A}(1-p)}$ and $\mathbf{B}_{NS}^{(M_{A},p)}=2^{M_{A}}$.
Notably, there is no violation of the no-signaling principle by setting
$\left\vert \left\langle \mathbf{B}_{l}\right\rangle _{NS}\right\vert =1$ with
unbiased local outcomes. In the following, the angle parameters in the
measurement assignment are set as $\frac{\pi}{4}$ for achieving the quantum
upper bounds.

Next we explore the star-shaped fully-quantum network $\mathcal{N}%
_{Q}^{(n,M_{A}+1)}$ with $e_{i}^{(Q)}$ emitting the $N_{i}$-qubit state
$\left\vert GHZ_{_{N_{i}}}\right\rangle $ stabilized by the stabilizing
operators $\pm O_{(i,1)}O_{(i,2)}\cdots O_{(i,N_{i})}$, where $O_{(i,j)}%
\in\{X_{(i,j)},Y_{(i,j)}\}$ for all $j=1,2,..,N_{i}$. $N_{i}-2\left\lfloor
\frac{N_{i}}{2}\right\rfloor $ of these $N_{i}$ Pauli observables, $O_{(i,1)}%
$, $O_{(i,2)}$,$\cdots$, $O_{(i,N_{i})}$, are $X$ and the other $2\left\lfloor
\frac{N_{i}}{2}\right\rfloor $ Pauli observables are $Y$. Here the qubit
$(i,k)$ emitted from $e_{i}^{(Q)}$ is sent to the observer Alice$_{i,k}$ if
$1\leq k\leq$ $K_{i}$, and sent to Bob if $K_{i}+1\leq k\leq$ $N_{i}$.
Regarding the state $\left\vert GHZ_{_{N_{i}}}\right\rangle $, there are
$2^{N_{i}-1}$ legal $N_{i}$-bit strings $l^{(i)}=l_{1}^{(i)}l_{2}^{(i)}\cdots
l_{N_{i}}^{(i)}$ such that the associated operator $g_{l^{(i)}}=$
$\pm(X\overline{^{l_{1}^{(i)}}}$ $Y^{l_{1}^{(i)}})\otimes$ $(X\overline
{^{l_{2}^{(i)}}}$ $Y^{l_{2}^{(i)}})\cdots\otimes$ $(X\overline{^{l_{K_{i}%
}^{(i)}}}$ $Y^{_{K_{i}}^{(i)}})\otimes O_{B,l^{(i)}}$ stabilizes $\left\vert
GHZ_{N_{i}}\right\rangle =\left\vert GHZ_{N_{i}}\right\rangle $, where
$O_{B,l^{(i)}}=%
%TCIMACRO{\dprod \nolimits_{i=K_{i}+1}^{n}}%
%BeginExpansion
{\displaystyle\prod\nolimits_{i=K_{i}+1}^{n}}
%EndExpansion
(X\overline{^{l_{K_{i}+1}^{(i)}}}$ $Y^{l_{K_{i}+1}^{(i)}})\otimes$
$(X\overline{^{l_{K_{i}+2}^{(i)}}}$ $Y^{l_{K_{i}+2}^{(i)}})\cdots\otimes$
$(X\overline{^{l_{N_{i}}^{(i)}}}$ $Y^{_{N_{i}}^{(i)}})$ and the pairty value $%
%TCIMACRO{\dsum \nolimits_{k=1}^{N_{i}}}%
%BeginExpansion
{\displaystyle\sum\nolimits_{k=1}^{N_{i}}}
%EndExpansion
l_{k}^{(i)}\operatorname{mod}2$ is a constant. Similarly, as for the observer
Alice$_{i,k}$, the type-$\mathcal{A}$ measurement assignment
reads$\ \ X\rightarrow\frac{(A_{0}^{(i,k)}+A_{1}^{(i,k)})}{\sqrt{2}}$ and
$Y\rightarrow\frac{(A_{0}^{(i,k)}-A_{1}^{(i,k)})}{\sqrt{2}}$ $\forall k=1$,
$2$,.., $K_{i}$; as for the observer Bob, the type-$\mathcal{B}$ measurement
assignment reads $O_{B,l^{(i)}}\rightarrow B_{l^{(i)}}$. The Bell operators
associated with $g_{l^{(i)}}$ reads $\mathbf{B}_{l^{(i)}}=\pm(%
%TCIMACRO{\dprod \nolimits_{k=1}^{K_{i}}}%
%BeginExpansion
{\displaystyle\prod\nolimits_{k=1}^{K_{i}}}
%EndExpansion
\frac{(A_{0}^{(i,k)}+(-1)l_{k}^{(i)}A_{1}^{(i,k)})}{2})B_{l^{(i)}}$, and the
Bell ineqaulity tailored for $\left\vert GHZ_{N_{i}}\right\rangle $ is
$\left\langle
%TCIMACRO{\dsum \nolimits_{l_{1}^{(i)},...,l_{K_{i}}^{(i)}=0}^{1}}%
%BeginExpansion
{\displaystyle\sum\nolimits_{l_{1}^{(i)},...,l_{K_{i}}^{(i)}=0}^{1}}
%EndExpansion
\mathbf{B}_{l^{(i)}}\right\rangle \overset{C}{\leq}1$ with the quanutm upper
bound being $2^{\frac{K_{i}}{2}}$. Denote the $M_{A}$-bit string
$l=l_{1}^{(1)}l_{2}^{(1)}\cdots l_{K_{1}}^{(1)}||l_{1}^{(2)}l_{2}^{(2)}\cdots
l_{K_{2}}^{(2)}|...||l_{1}^{(n)}l_{2}^{(n)}\cdots l_{K_{n}}^{(n)}$ with
$M_{A}=$ $%
%TCIMACRO{\dsum \nolimits_{i=1}^{k}}%
%BeginExpansion
{\displaystyle\sum\nolimits_{i=1}^{k}}
%EndExpansion
K_{i}$, and the associated stabilizing operator $g_{l}=%
%TCIMACRO{\dprod \nolimits_{i=1}^{n}}%
%BeginExpansion
{\displaystyle\prod\nolimits_{i=1}^{n}}
%EndExpansion
g_{l^{(i)}}$ such that $g_{l}\left\vert \Psi\right\rangle =\left\vert
\Psi\right\rangle =$ $%
%TCIMACRO{\dprod \nolimits_{i=1}^{n}}%
%BeginExpansion
{\displaystyle\prod\nolimits_{i=1}^{n}}
%EndExpansion
\left\vert GHZ_{_{N_{i}}}\right\rangle $. As a result, we can construct the
Bell operator$\ \mathbf{B}_{l}=%
%TCIMACRO{\dprod \nolimits_{i=1}^{n}}%
%BeginExpansion
{\displaystyle\prod\nolimits_{i=1}^{n}}
%EndExpansion
$ $\mathbf{B}_{l^{(i)}}$. It is easy to verify that the Bell inequality
tailored for the distributed quantum state $\left\vert \Psi\right\rangle $ is
$%
%TCIMACRO{\dsum \nolimits_{l}}%
%BeginExpansion
{\displaystyle\sum\nolimits_{l}}
%EndExpansion
\left\vert \left\langle \mathbf{B}_{l}\right\rangle _{C}\right\vert ^{p}%
\leq2^{M_{A}(1-p)}=\mathbf{B}_{C}^{(M_{A},p)}$, where and the quantum and
no-signaling upper bounds $\mathbf{B}_{Q}^{(M_{A},p)}$ and $\mathbf{B}%
_{NS}^{(M_{A},p)}$ are also equal to $2^{M_{A}(1-\frac{p}{2})}$ and $2^{M_{A}%
}$, respectively.

Finally, we explore alternative linear Bell inequalities for the star-shaped
network $\mathcal{N}_{Q}^{(2s+1,2s+2)}$ with each quantum source emitting the
two-qubit Bell states. The qubits $(1,1)$, $(2,1),\cdots,$and $(2s+1,1)$ are
sent to $\mathcal{A}_{1}$ that performs the type-$A$ measurements on these
$(2s+1)$ qubits. On the other hand, the qubit $(j,2)$ is sent to
\ $\mathcal{B}_{j}$ that performs the type-$B$ measurement, $j=1,...,2s+1$.
Given the\ $(2s+1)$ -bit string $l=l_{1}...l_{2s+1}$ with the even parity
constraint $%
%TCIMACRO{\dsum \nolimits_{i=1}^{2s+1}}%
%BeginExpansion
{\displaystyle\sum\nolimits_{i=1}^{2s+1}}
%EndExpansion
l_{i}\operatorname{mod}2=0$, denote the associated stabilizing operator is
$g_{l}=%
%TCIMACRO{\dprod \nolimits_{i=1}^{2s+1}}%
%BeginExpansion
{\displaystyle\prod\nolimits_{i=1}^{2s+1}}
%EndExpansion
(Z_{(i,1)}Z_{(i,2)})^{\overline{l_{i}}}(X_{(i,1)}X_{(i,2)})^{l_{i}}$ and the
complementary $(2s+1)$ -bit string $\overline{l}=\overline{l}_{1}%
...\overline{l}_{2s+1}$, where $\overline{l_{i}}=l_{i}+1\operatorname{mod}2$.
Denote $B_{l}=%
%TCIMACRO{\dprod \nolimits_{i=1}^{2s+1}}%
%BeginExpansion
{\displaystyle\prod\nolimits_{i=1}^{2s+1}}
%EndExpansion
(Z_{(i,1)}^{\overline{l_{i}}}X_{(i,1)}^{l_{i}})$ and, as for the type-$A$
measurement, assign $B_{l}\rightarrow\frac{(A_{l}^{(1)}+A_{\overline{l}}%
^{(1)})}{\sqrt{2}}$ and $B_{\overline{l}}\rightarrow\frac{(A_{l}%
^{(1)}-A_{\overline{l}}^{(1)})}{\sqrt{2}}$. \bigskip Notably, we have
$\left\{  B_{l},B_{\overline{l}}\right\}  =\left\{  A_{l},A_{\overline{l}%
}\right\}  =0$, and $A_{l}^{2}=A_{\overline{l}}^{2}=I_{2^{2s+1}}$, where
$I_{2^{2s+1}}$ is $2^{2s+1}\times2^{2s+1}$ unit matrix. Here we introduce the
following lemma.

\textit{Lemma }Let $\widetilde{X}$ and $\widetilde{Z}$ be Hermitian operators
acting on $H$ such that dim $H=d=2d^{\prime}<\infty$ . If $(a)$ $\widetilde
{X}^{2}$ $=\widetilde{Z}^{2}$ $=$ $I_{H}$ and $(b)$ $\left\{  \widetilde
{X},\widetilde{Z}\right\}  =0$, we have $H=C^{2}\otimes C^{d^{\prime}}$ \ and
there exists a local unitary operation $U$ such that $U\widetilde{X}%
U^{\dagger}=X\otimes I_{d^{\prime}\text{ }}$and that $U\widetilde{Z}%
U^{\dagger}=Z\otimes I_{d^{\prime}\text{ }}$.

The reader can refer to \cite{s1} for the detailed proof. In this way, given
an arbitrary $(2s+1)$ -bit string $l$, we can regard the observable pair
$(B_{l}$, $B_{\overline{l}})$ as $(\widetilde{Z}$, $\widetilde{X})$ on a
virtual qubit, and hence we have $M_{A}=1$ according to the lemma. As the
type-$B$ measurement, assign $Z_{(j,2)}\rightarrow B_{0}^{(j)}$ and
$X_{(j,2)}\rightarrow B_{1}^{(j)}$. In this case, $M_{A}=1$, $M_{B}=2s+1$. As
a result, given the\ $(2s+1)$ -bit string $l=l_{1}...l_{2s+1}$ with even
parity, the associated segmented Bell operator $\mathbf{B}_{l}=\frac
{(A_{l}^{(1)}+A_{\overline{l}}^{(1)})}{\sqrt{2}}%
%TCIMACRO{\dprod \nolimits_{i=1}^{2s+1}}%
%BeginExpansion
{\displaystyle\prod\nolimits_{i=1}^{2s+1}}
%EndExpansion
(B_{0}^{(j)})^{\overline{l_{i}}}(B_{1}^{(j)})^{l_{i}}$ and $\mathbf{B}%
_{\overline{l}}=\frac{(A_{l}^{(1)}-A_{\overline{l}}^{(1)})}{\sqrt{2}}%
%TCIMACRO{\dprod \nolimits_{i=1}^{2s+1}}%
%BeginExpansion
{\displaystyle\prod\nolimits_{i=1}^{2s+1}}
%EndExpansion
(B_{0}^{(j)})^{\overline{l_{i}}}(B_{1}^{(j)})^{l_{i}}$, the proposed Bell
inequality reads
\[
\left\vert \left\langle \mathbf{B}_{l}\right\rangle _{C}\right\vert
^{p}+\left\vert \left\langle \mathbf{B}_{\overline{l}}\right\rangle
_{C}\right\vert ^{p}\leq2^{1-p}%
\]
with the quantum upper bound $2^{1-\frac{p}{2}}$. Some remarks are in order.
Note that there are $2^{2s}$ $(2s+1)$-bit strings with even parity, and hence
there are $2^{2s}$ proposed Bell inequalities tailored for the quantum state
$\left\vert Bell\right\rangle ^{\otimes n}$ distributed in the star-shaped
network $\mathcal{N}_{Q}^{(2s+1,2s+2)}$. On the other hand, the proposed Bell
inequality still holds even if we set $\mathcal{B}_{1}=\mathcal{B}%
_{2}=\mathcal{B}_{2s+1}=\mathcal{B}$ that measures the joint observable $%
%TCIMACRO{\dprod \nolimits_{i=1}^{2s+1}}%
%BeginExpansion
{\displaystyle\prod\nolimits_{i=1}^{2s+1}}
%EndExpansion
(B_{0}^{(j)})^{\overline{l_{i}}}(B_{1}^{(j)})^{l_{i}}$ or $%
%TCIMACRO{\dprod \nolimits_{i=1}^{2s+1}}%
%BeginExpansion
{\displaystyle\prod\nolimits_{i=1}^{2s+1}}
%EndExpansion
(B_{0}^{(j)})^{\overline{l_{i}}}(B_{1}^{(j)})^{l_{i}}$ on the $(2s+1)$ qubits.
Finally, it is straightforward to verify that the quantum upper bound of
$\left\vert \left\langle \mathbf{B}_{l}\right\rangle _{Q}\right\vert
^{p}+\left\vert \left\langle \mathbf{B}_{\overline{l}}\right\rangle
_{Q}\right\vert ^{p}$ is $2^{1-\frac{p}{2}}$.

\end{document}